\documentclass[12pt]{article}

\usepackage{amsthm}
\usepackage{amsmath}
\usepackage{natbib}
\usepackage[colorlinks,citecolor=blue,urlcolor=blue,filecolor=blue,backref=page]{hyperref}
\usepackage{graphicx}

\usepackage{enumerate}
\usepackage{multirow}
\usepackage{verbatim}
\usepackage[font=footnotesize]{subcaption}
\usepackage{amsfonts}
\usepackage{amssymb}
\usepackage{color,soul}
\usepackage{hyperref}

\usepackage{enumitem, booktabs,rotating}
\usepackage{mathrsfs}
\usepackage{url}

\newtheorem{theorem}{Theorem}[section]

\newcommand{\blind}{1}

\begin{document}

\def\spacingset#1{\renewcommand{\baselinestretch}%
{#1}\small\normalsize} \spacingset{1}


\if1\blind
{
  \title{\bf A Bayesian Nonparametric Approach for Identifying Differentially Abundant Taxa in Multigroup Microbiome Data with Covariates}
   \author{Archie Sachdeva
\\
   Bristol Myers Squibb, Princeton, NJ\\
    and \\
    Somnath Datta\\
    Department of Biostatistics, University of Florida\\
    and \\
     Subharup Guha\\
    Department of Biostatistics, University of Florida}
  \maketitle
} \fi

\if0\blind
{
  \bigskip
  \bigskip
  \bigskip
  \begin{center}
    {\LARGE\bf Causally Interpretable Meta-Analysis of Observational Studies}
\end{center}
  \medskip
} \fi

\bigskip
\begin{abstract}
Scientific studies in the last two decades have established the central role of the microbiome in disease and health. Differential abundance analysis seeks to identify microbial taxa associated with  sample groups defined by a factor  such as disease subtype, geographical region, or environmental condition. The results, in turn, help clinical practitioners and researchers diagnose disease and develop  treatments more effectively. However, microbiome data analysis is uniquely challenging due to high-dimensionality, sparsity, compositionally, and collinearity. There is a critical need for unified statistical approaches for differential analysis in the presence of covariates. We develop a zero-inflated Bayesian nonparametric (ZIBNP) methodology that meets these multipronged challenges. The proposed technique flexibly adapts to the unique data characteristics, casts the  high proportion of zeros in a censoring framework, and mitigates high-dimensionality and collinearity by utilizing the dimension-reducing property of the semiparametric Chinese restaurant process. Additionally, the ZIBNP approach relates the microbiome sampling depths to inferential precision while accommodating the compositional nature of microbiome data. Through simulation studies and analyses of the CAnine Microbiome during Parasitism (CAMP) and Global Gut microbiome datasets, we demonstrate the accuracy of ZIBNP compared to established  methods for differential abundance analysis in the presence of covariates.
\end{abstract}

\noindent%
\vfill

\spacingset{1.2} 

\section{Introduction}\label{S:introduction}

Numerous scientific studies have  established the influence of the 
 microbiome  on  health and disease \citep{WADE2013137,Shreiner2015,rautava_2016}. 
Giant strides in next-generation sequencing technologies  offer  a glimpse into the  microbiome present in our bodies by facilitating the simultaneous identification and quantification of  taxa abundances  \citep{Tang2019}, where a \textit{taxon}   represents any profiled set of microbiome features such as operational taxonomic units (OTUs), amplicon sequence variants (ASVs), or metagenomic species.  \textit{Differential abundance  analysis}, or the identification of  microbial  taxa  associated with population groups, plays a key role in disease diagnoses and the development of novel treatments \citep{cappellato2022investigating}. Informally, a taxon is   \textit{not} differentially abundant if, after adjusting for the covariates, every pair of samples belonging to different groups have similar normalized abundances for that taxon; otherwise, the taxon is differentially abundant. This is especially challenging because microbiome abundance data are  high-dimensional,  sparse, and compositional,  necessitating sophisticated statistical methods \cite[e.g.,][]{Weiss2017}. It is also essential to  adjust for  individual-specific attributes in  differential abundance  analyses \citep{Vujkovic-Cvijin2020}; for example, when the groups are disease subtypes, covariates such as  diet and lifestyle that  influence   microbial composition  may    be highly associated with the grouping variable.   

Microbiome samples are  sequenced using one of two  processes: amplicon sequencing and random shotgun sequencing. Shotgun sequencing is restricted to  highly variable regions of the 16S rRNA gene \citep{Kembel_16S2012},  whereas amplicon  sequencing  assesses genome-wide genetic variation. Shotgun  sequencing  has  better species-level resolution and  ability to detect novel viruses  than amplicon  sequencing \citep{BIBBY2013275,Relman2013}, but  requires considerably larger  volumes \citep{Sharpton_2014}. We refer the reader to \cite{Bharti2019}   for a detailed review of   sequencing technologies and bioinformatics pipelines  for analyzing
raw microbiome  data  \citep{Xia2018_ch1}.  
Subsequently,  taxonomic profile generation \citep{Samuel_2021,Bharti2019} maps  the microbiome sequences  to known reference databases and produces a high-dimensional vector of abundance counts for each  sample \citep{Schloss2011}. Each  element of this vector corresponds to a taxon. The abundance counts of the  samples are arranged in a matrix called the \textit{abundance table}; each   element represents the measured  abundance  for a  taxon (column) in a  sample (row). The number of taxa far exceeds the number of  samples, resulting in a ``short and wide'' abundance table.

The term ``abundance counts'' is somewhat misleading because
high-throughput  sequencing technologies do not  measure the actual  counts. Instead, the  counts are measured relative to the total number of reads,  commonly referred to as the \textit{library size} or \textit{sampling depth},   related to  measurement precision \citep[e.g.,][]{Weiss2017}. After normalization, the sample-specific   measurements (i.e., abundance matrix rows) are intrinsically  \textit{compositional}  and  represent random points on a high-dimensional simplex \citep{aitchison_2008}. Furthermore, due to technical variations of the sequencing procedure, the library sizes  vary from sample to sample  \citep{McKnight2019,Weiss2017}. 
 Microbiome data are also highly \textit{sparse} and  comprise  $10\%$ - $70\%$ zeros. Some zeros 
are caused by  individual  traits; for instance, low-fiber diets are known to deplete certain gut bacteria  \citep{diet2020}. However, most zeros occur due to sequencing process errors or  relatively small sampling depths. In the literature, these three types of zeros are respectively referred to as biological, sampling, and technical zeros    \citep{SILVERMAN20202789, Kaul2017,Weiss2017}.

An  overview of state-of-the-art, and predominantly frequentist, methods for  differential abundance analysis can be found in  \cite{Wallen2021} and \cite{dattaGuha2021}. These  methods are   broadly classified as \textit{compositional} or \textit{count-based}  \citep{Liu2021,Nearing2022}. 
Prominent among  compositional approaches   are ALDEx2 \citep{Fernandes2014} and ANCOM-II \citep{Kaul2017}, both of which use Wilcoxon rank-sum tests to detect the differential taxa of the groups. ALDEx2 relies on the centered log-ratio (clr) transformation and Dirichlet-multinomial model;  ANCOM-II uses the additive log-ratio (alr) transformation. 
 By contrast,  count-based approaches utilize  the sampling depths to inform inferential precision.
Zero-inflated count approaches that attempt to account for sparsity  include metagenomeseq \citep{Paulson2013}, which  relies on a zero-inflated log-normal model, and \cite{Risso2018} and \cite{Xia2018_bookCH_ZI}, which  account for overdispersion using negative binomial models. ANCOMBC \citep{Lin2020} uses a log-linear regression framework  with a bias-corrected random intercept. For two-group comparisons, MaAsLin2 \citep{maaslin2_2021} uses  GLMs along with different normalization  and  distributional options. The methods DESeq2 \citep{Love2014} and corncob \citep{Martin2020} respectively utilize negative binomial and beta-binomial distributions and   assume  marginal distributions  for the featurewise counts.  However, no    statistical method addresses  the disparate challenges of microbiome data or performs reliably with different microbiome datasets~\citep{Weiss2017}.

We propose a coherent statistical framework called the zero-inflated Bayesian nonparametric (ZIBNP) model that detects   differentially abundant taxa of multiple groups of  samples while effectively confronting the multipronged  challenges of  high-dimensionality, compositionality, collinearity, sparsity, and covariate confounding. As its name suggests, the hierarchical model comprises a two-component mixture model for the abundance counts with one mixture component being a point mass at zero and the other component comprising a flexible Bayesian nonparametric distribution whose posterior  adapts to the  dataset characteristics. Technical and sampling zeros, introduced earlier, are cast in a censoring framework and   inferred a posteriori. ZIBNP is a hybrid of compositional and count-based methods because it relates the   sample covariates and   library sizes to inferential precision while also  accommodating the  compositional aspect of microbiome data.  High-dimensionality and collinearity (specifically,  small $n$, large $p$ issues) are mitigated  by allocating the large number of taxa to fewer latent clusters defined by shared  relative abundance patterns using the semiparametric Chinese restaurant process \citep{MullerMitra2013,Lijoi_Prunster_2010}. 
We apply the ZIBNP technique  to infer the differentially abundant  taxa in two publicly available  microbiome datasets: \textit{(i)} The CAnine Microbiome during Parasitism (CAMP) study \citep{mdb} explores the impact of natural parasite infection on the gut microbiome of   infected and uninfected domesticated dogs as the two groups; and \textit{(ii)} Global Gut microbiome data \citep{Yatsunenko2012} on humans  residing in three geographical regions  as the groups.

The paper is organized as follows. Section \ref{S:model} introduces  the ZIBNP model. Section~\ref{S:BNP} develops the nonparametric mixture component   for mitigating high-dimensionality and collinearity,  incorporating covariate effects, and detecting  differential taxa. Section \ref{S:technical zeros} addresses the challenges of sparsity by fostering a censoring framework for  non-biological zeros. Section \ref{S:inference} outlines the posterior inference procedure.  Section \ref{S:simulation} uses artificial datasets to demonstrate   the high accuracy achieved by ZIBNP relative to  existing   approaches that  adjust for sample covariates.   Section \ref{sec:dataAnalysis}  analyzes the  motivating microbiome datasets using the proposed ZIBNP method.  Section \ref{S:discussion} concludes with a brief discussion.

\section{A  Bayesian hierarchical approach} \label{S:model}

We foster a Bayesian framework capable of  inferring  the complex relationships between   microbial taxa and groups defined by a    factor  such as disease status, disease subtype, or geographical region, and 
 identifying the differential abundant (DA)  taxa associated with the groups. The taxa-group relationships  may be confounded by  covariates  producing spurious  associations  \citep{Vujkovic-Cvijin2020}.
 Of primary interest are latent \textit{differential status variables} $\tilde{h}_j$, defined as  $\tilde{h}_j=2$ ($\tilde{h}_j=1$) if taxon $j$ is (not)  DA. The set of DA taxa  is  
 $\mathscr{D} = \left\{j: \tilde{h}_{j}=2,  \,\, j=1,\ldots,p\right\}$. 
 
For $n$ samples and $p$ taxa, the  data are arranged in  an $n \times p$ abundance matrix with each matrix element representing the observed count or abundance of a taxon (matrix column) in a  study sample or subject (matrix row). Usually, there are far fewer samples than taxa, i.e., $n << p$. We write the taxa abundance matrix as ${\mathbf{Z}} = (({Z}_{ij}))$, where $Z_{ij}$ is the number of reads of taxon $j$ in sample $i$, and $\mathbf{Z}_i=(Z_{i1},\ldots,Z_{ip})'$ is the vector of taxa abundances for sample $i$. Suppose there are $T$ covariates, $\mathbf{X}_i$, and the $n$ by $T$ covariate matrix is denoted by $\mathbf{X}=((X_{il}))$. With ${L_i}$ representing the observed sequencing depth or library size, the \textit{relative abundance} vector of the $i$th subject,  denoted by $\hat{\mathbf{q}}_i=(\hat{q}_{i1},\ldots,\hat{q}_{ip})'$, equals   $\mathbf{Z}_i/L_i$.
 For $K\ge 2$ groups,
the group membership of the $i$th sample is denoted by $k_i$, and there are $n_k$ samples belonging to the $k$th group, $k=1,\ldots,K$.     
 Now suppose there are two samples, $i_1$ and $i_2$, belonging to different groups ($k_{i_1}  \neq k_{i_2}$) but having  similar  covariates ($\mathbf{X}_{i_1} \approx \mathbf{X}_{i_2}$). Then, by the intuitive characterization of differential statuses in Section \ref{S:introduction}, a taxon $j$ is non-DA if $\hat{q}_{{i_1}j}\approx \hat{q}_{{i_2}j}$. However, if $\hat{q}_{{i_1}j}$ very different from $\hat{q}_{{i_2}j}$, then taxon $j$ is DA.

As discussed in Section \ref{S:introduction}, 
\textit{biological zeros} refer to the complete absence of a taxon in an entire (say, $k$th) group  and manifest as a  matrix column of zeros for all $n_k$  subjects. These taxa are easily identified as DA and removed from the dataset in a pre-processing step.  On the other hand, \textit{technical zeros} are caused by   processing or sequencing batch effects and \textit{sampling zeros}   are randomly missing taxa   in specific samples. Distinguishing and appropriately accounting for  these ``non-biological'' zeros in differential analysis requires more sophisticated  techniques.     
  
The proposed ZIBNP model postulates a two-component mixture for each element of $\mathbf{Z}$, namely, a point mass at zero representing  technical zeros and a  Bayesian nonparametric (BNP) model, denoted by $\mathscr{F}$, under which sampling zeros may  stochastically occur. 
More formally, with $\mathscr{F}_{ij}$ denoting the marginal distribution of  $Z_{ij}$ under BNP model $\mathscr{F}$ and $I_{\{0\}}$ representing a point mass at~0:

    \begin{equation}
    Z_{ij} \mid r_{ij}, \mathscr{F}_{ij} \overset{\text{indep}}{\sim} r_{ij} I_{\{0\}} + (1-r_{ij}) \mathscr{F}_{ij},\quad i=1,\ldots,n;\,\, j=1,\ldots,p,\label{eq:mixture}
    \end{equation}
where $r_{ij}$ is the probability of a technical zero in sample $i$ and taxon $j$. Let $\mathbf{R}=((r_{ij}))$ denote the $n \times p$ matrix of  technical zero probabilities. The first mixture component represents technical zeros and the second  component includes sampling zeros.  Section~\ref{S:BNP} develops  BNP model  $\mathscr{F}$.  Section~\ref{S:technical zeros} describes the model for  technical zeros.

\subsection{Bayesian nonparametric (BNP) model}\label{S:BNP}

The BNP model $\mathscr{F}$ assumes the following  distribution for the taxa abundances:
\begin{equation}
        {\mathbf{Z}}_i \mid \mathbf{q}_i \overset{\text{indep}}{\sim} \text{Multinomial}({L_i}, \mathbf{q}_i), \quad i=1,\ldots, n, \label{eq:Zi}
\end{equation}
where probability $p$-tuple $\mathbf{q}_i=(q_{i1},\ldots,q_{ip})'$ is a less noisy version of  relative abundance vector $\hat{\mathbf{q}}_i=(\hat{q}_{i1},\ldots,\hat{q}_{ip})'$  and forms the $i$th row of an $n$ by $p$  matrix, $\mathbf{Q}=((q_{ij}))$. Let $\tilde{\mathbf{q}}_j$ denote the $j$th matrix column, so that $\mathbf{Q}=\bigl[\tilde{\mathbf{q}}_1,\ldots,\tilde{\mathbf{q}}_p\bigr]$. The conditional probability of a sampling zero  is   
\begin{equation}P\bigl[Z_{ij} =0\mid \mathbf{Q}, \mathscr{F}\bigr] = (1-q_{ij})^{L_i}.\label{eq:sampling0}
\end{equation}

\paragraph*{Dimension reduction by unsupervised clustering} \quad One of the challenges of 
microbiome data analysis is  the  ``small $n$, large $p$'' setting causes severe collinearity in the relative abundance  matrix  and  gives inefficient inferences. BNP model $\mathscr{F}$ resolves these issues by  utilizing lower-dimensional matrix structure. 
Specifically,  unsupervised clustering of the $p$ taxa  allocates them to  $H$  latent clusters, where $H$ is a priori unknown. Each  cluster is  characterized by a  $n$-variate  \textit{motif} delineating the   relative  abundance pattern shared by all taxa  belonging to the cluster. Dimension reduction is achieved when  $H$ is  less than~$p$.  

From a mathematical standpoint, latent clusters are unavoidable in small $n$, large $p$ matrices because 
the matrix rank cannot exceed $n$, and this produces  redundancies in the large number of matrix columns. From a scientific standpoint as well,   biomarkers belonging to  common  phylogenetic groups or functional pathways tend to  be highly correlated  \citep[e.g.,][]{lee2020cococonet}. The phenomenon of   biomarker clusters having   similar patterns in high-dimensional genomic, epigenomic, and transcriptomic datasets is well-documented and   has been  utilized  to achieve dimension reduction  \citep[e.g.,][]{Medvedovic_etal_2004,Kim_etal_2006, Dunson_etal_2008,Dunson_Park_2008,guha2022predicting,gu2023nonparametric}.

 More formally, define taxon-to-cluster  \textit{mapping variables}, $c_1,\ldots,c_p$, with the random event $\{c_j=u\}$ indicating  that the $j${th} taxon is allocated to the $u${th} latent cluster,  $u=1,\ldots,H$. 
  Let $m_u = \sum_{j=1}^{p}I(c_j =u)$ be the  number of taxa in the $u$th latent cluster.
 Mapping vector $\boldsymbol{c}=(c_1,\ldots,c_p)$ is  given a  Chinese restaurant process (CRP) prior with a positive precision or mass parameter $\alpha$ \citep{MullerMitra2013}. With $[\cdot]$ representing densities with respect to a dominating measure, we have 
\begin{equation}\bigl[\boldsymbol{c}\mid \alpha\bigr] = \frac{\Gamma(\alpha) \,\alpha^H}{\Gamma(\alpha + p)} \prod_{u=1}^{H} \Gamma(m_u), \quad \boldsymbol{c} \in \mathscr{P}_p, \label{eq:CRP}
\end{equation}
 where  $\mathscr{P}_p$  is the set of all partitions of  $p$ taxa into one or more latent clusters. 
  CRPs  achieve dimension reduction in the large number of taxa because the random number of clusters, $H$, is asymptotically equivalent to $\alpha \log(p)$ as $p \to \infty$ \citep{Lijoi_Prunster_2010}, so that typically $H\ll p$.

\paragraph{Cluster motifs} \quad  Motif $\mathbf{q}_u^*=(q_{1u}^*,\ldots,q_{nu}^*)'$   embodies the ``signal'' underlying the   relative abundances  of the $u$th cluster's member taxa. In other words,    the $H$ motifs  represent the    across-sample pattern shared by all taxa belonging to a latent cluster. 
Expression~(\ref{eq:Zi}) establishes that     relative abundance vectors $\hat{\mathbf{q}}_1,\ldots,\hat{\mathbf{q}}_p$ are just  noisy versions of the corresponding matrix $\mathbf{Q}$ columns, $\tilde{\mathbf{q}}_1,\ldots,\tilde{\mathbf{q}}_p$. This suggests that  the  matrix columns $\mathbf{Q}$ are  identical to their motifs: 
\begin{equation}
    \tilde{\mathbf{q}}_j=\mathbf{q}_{c_j}^*, \quad\text{for  taxa $j=1,\ldots,p$}. \label{eq:qtilde}
\end{equation}
Because the measured taxa abundances in (\ref{eq:Zi}) are random, the taxa belonging to a cluster have similar, although not  necessarily identical,  relative abundances.

Conditional on the mapping vector $\boldsymbol{c}=(c_1, \ldots, c_p)$, matrix
  $\mathbf{Q}$  is  fully determined by the lower-dimensional matrix $\mathbf{Q}^*=((q_{iu}^{*}))$ of dimension $n$ by $H$, where $H \ll p$.   Additionally, 
 since  $\mathbf{Q}$ is row-stochastic, each row of matrix $\mathbf{Q}^*$  satisies
 \begin{equation}
     \sum_{u=1}^{H} m_u q_{iu}^* =1, \quad  i=1,\ldots,n, \label{eq:constraint}
 \end{equation}  
 implying that the cluster motifs  are linearly dependent. 
 Furthermore, applying equation~(\ref{eq:qtilde}) and the informal characterization of differential taxa in Section \ref{S:introduction}, we find
 that  all taxa in a latent cluster  must have identical differential statuses. That is,  a cluster is collectively   DA or non-DA. Consequently, analogously to the $p$ taxa differential status variables, we can   define  \textit{cluster} differential status variables $h_1,\ldots,h_H$ and describe the set of DA taxa equivalently as
\begin{equation}
    \mathscr{D} = \left\{j: h_{c_j}=2,  \,\, j=1,\ldots,p\right\}. \label{eq:D}
\end{equation}

\paragraph*{Incorporating covariates} \quad To model the regression relationships of the cluster motifs, we begin by selecting     
a non-DA \textit{reference cluster}. 
Strategies for choosing this special non-DA cluster 
are discussed in the sequel. Because the CRP  is symmetric with respect to the cluster labels,  the reference 
cluster is labeled $u=1$ without loss of generality. We  assume  
\begin{align}
     \eta_{iu} =\log(\frac{q_{iu}^*}{q_{i1}^*}) \stackrel{\text{indep}}\sim N\biggl(\beta_{0k_iu} + \sum_{l=1}^Tx_{il} \beta_{lk_iu}, \,\, \sigma_e^2\biggr), \label{eq:eta}\\  \text{for $i=1,\ldots,n$, and $u>1$.}  \notag
\end{align}
The  $q_{iu}^*$ are  identifiable due to constraint~(\ref{eq:constraint}) and matrix $\mathbf{Q}$  is  then available from (\ref{eq:qtilde}). We assume that  $\sigma_e^2$ is small enough that most of the variability in the $\eta_{iu}$'s is captured by the regression means. As a result, in the (unrealistic) situation in which there are no  covariates  (i.e., $x_{il}\equiv 0$), the latent elements $q_{iu}^*$ of all $n_k$ samples in any  group $k$ are approximately equal. 

\paragraph*{Differential statuses of clusters and taxa} \quad 
By design, the reference cluster (taxon) is  non-DA, and the differential status variable for the cluster is $h_1=1$.
For clusters~$u>1$, 
since the regression vectors $\boldsymbol{\beta}_{1u},\ldots, \boldsymbol{\beta}_{Ku}$ adjust for the covariates, the cluster differential statuses are  available as
\begin{align}
h_u=
\begin{cases}
1  \qquad\text{if } \boldsymbol{\beta}_{1u}=\dots=\boldsymbol{\beta}_{Ku}, \\
2 \qquad \text{otherwise}.
\end{cases}
\label{eq:DA}
\end{align}
In other words, a non-reference cluster is non-DA if and only if its $K$ regression vectors   are  identical.   Further, the differential statuses of the  taxa  are  identical as their parent clusters. Returning to a hypothetical  no-covariate dataset, for any non-DA cluster $u$, expressions (\ref{eq:eta}) and (\ref{eq:DA}) show that the latent elements $q_{iu}^*$ of all $n$ samples, irrespective of their groups, are approximately equal. In the presence of covariates,    the differential statuses  may  not be obvious from the sample patterns of  $q_{iu}^*$ or the noisier relative abundances, but can be inferred using  (\ref{eq:DA}).

Evidently, a reasonable prior must allow multiple regression vectors  to be  equal with positive  probability. An appropriate prior can be specified  as follows. Let $M$ be a positive integer,  $\mathbf{1}_M$ be the column vector of $M$ ones, and $\mathbf{X}^{\dag}=[\mathbf{1}_n: \mathbf{X}]$ be the  extended covariate matrix of dimension $n$ by $(T+1)$.
Let $\delta_{\boldsymbol{\mu}}$ denote a point mass at $\boldsymbol{\mu}$. For a parameter vector $\boldsymbol{\gamma}$ with positive elements,  $\text{Dir}_{M}(\boldsymbol{\gamma})$ represents a Dirichlet distribution in $\mathcal{R}^M$.  The  regression vectors
have the  $M$-component  mixture prior: 
\begin{align}
        \boldsymbol{\beta}_{ku}  &\stackrel{\text{i.i.d.}}\sim  \sum_{m=1}^{M} \pi_m \delta_{\boldsymbol{\mu}_m}, \quad \text{$k=1,\ldots,K$, and $u> 1$, where}\label{eq:beta}\\
            \boldsymbol{\pi} =(\pi_1,\ldots,\pi_M)' &\sim \text{Dir}_{M}\bigl(\frac{\alpha_0}{M}\mathbf{1}_M\bigr), \quad \alpha_0>0, \notag\\ 
            \boldsymbol{\mu}_m &\stackrel{\text{i.i.d.}}\sim N_{T+1}\bigl(\mathbf{0},\tau^2 ({\mathbf{X}^{\dag}}^T\mathbf{X}^{\dag})^{-1}\bigr), \quad m=1,\ldots,M,\label{eq:mum_prior}
\end{align}
and where $\tau^2$ follows an inverse-gamma hyperprior with parameters $a_{\tau}$ and $b_{\tau}$. The finite mixture specification  is key because it allows  ties among the  $\boldsymbol{\beta}_{ku}$. In simulation studies and data analyses, we have found that  $M$ equal to 5 or 6 gives satisfactory results.

\begin{figure}
    \centering
    \includegraphics[width=0.95\textwidth]{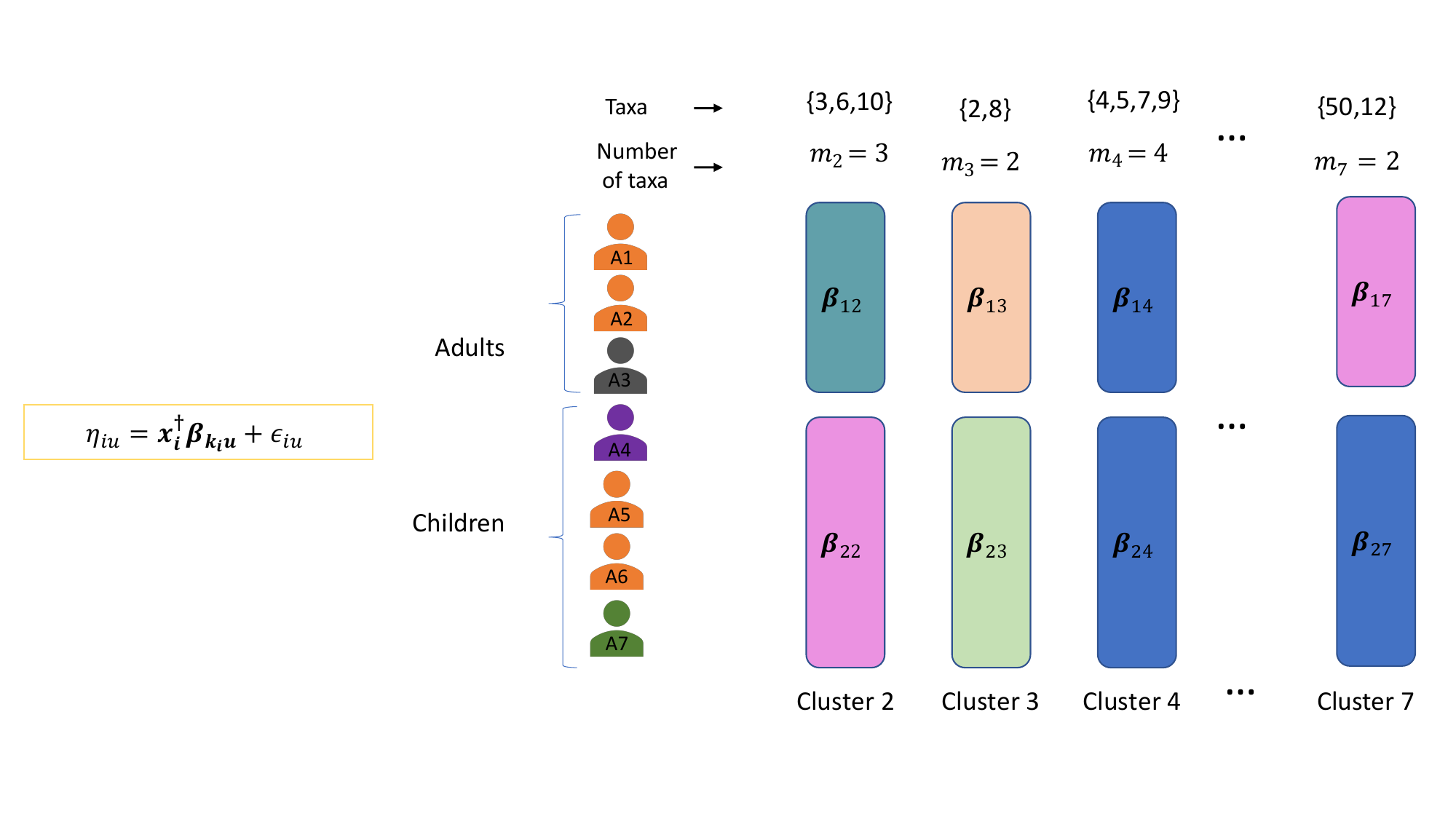}
    \caption{Cartoon illustration of the model-based procedure for calling the differential statuses of the taxa. There are $K=2$ subject groups  (adults and children), $p=50$ taxa,  multiple subject-specific covariates, and  $H=7$ latent clusters. The regression vectors for the non-reference clusters ($u>1$) are shown. The different colors identify the corresponding mixture  components  from which each $\boldsymbol{\beta}_{ku}$ is drawn. Applying criterion (\ref{eq:DA}), all taxa belonging to clusters 2, 3 and 7 are DA, and all taxa belonging to cluster~4 are non-DA.  See the text for  details.}
    \label{fig:toyDA}
\end{figure}

These ideas are demonstrated by a toy example in Figure \ref{fig:toyDA}, where we have $K=2$ groups of  subjects (i.e., adults and children), $p=50$ taxa, and several subject-specific covariates. The taxa are allocated to $H=7$ clusters. In the figure, row vector $\mathbf{X}^{\dag}_i$  denotes the $i$th row  of matrix $\mathbf{X}^{\dag}$ of length $(T+1)$ and $\epsilon_{iu} \stackrel{\text{i.i.d.}}\sim N(0, \sigma_e^2)$. The $K(H-1)=12$ vectors, each consisting of the $(T+1)$ regression coefficients for a group--non-reference cluster combination, arise from a finite mixture model with $M=5$ multivariate components. For $u>1$ (i.e., non-reference clusters), the colors represent the five-mixture model components  from which each $\boldsymbol{\beta}_{ku}$ vector is drawn  in (\ref{eq:beta}). For example, the  regression vectors $\boldsymbol{\beta}_{14}$,  $\boldsymbol{\beta}_{24}$, and $\boldsymbol{\beta}_{27}$ are   identical because they are  drawn from the same  mixture component (shown in blue). Applying  criterion (\ref{eq:DA}), cluster 4 is non-DA, and clusters 2, 3, and 7 are DA  in adults and children. All  taxa have the same differential status as their parent cluster. Specifically,  taxa 4, 5, 7, and 9,  which belong to cluster 4, are all non-DA; taxa 2 and 8, which  belong to cluster 3, are  both DA.

\paragraph{Choosing the reference cluster} \quad  
For correctly calling DA clusters using criterion (\ref{eq:DA}), it is important to choose a non-DA cluster as the reference. This can be seen by the following example. For simplicity, imagine a no-covariate dataset with $K=2$ groups,  $H=3$ latent clusters, and small $\sigma_e^2$ in (\ref{eq:eta}). Suppose we pick a  reference cluster ($u=1$) for which $q^*_{i_11}=0.5$ and $q^*_{i_21}=0.1$ for all samples $i_1$ and $i_2$ for which $k_{i_1}=1$ and $k_{i_2}=2$. That is, a DA cluster was mistakenly chosen to be the reference. Now, let cluster $u=2$ be actually non-DA with $q^*_{i2}=0.1$ for all $i$.  Since $\sigma_e^2$ is small, relation (\ref{eq:eta}) gives $\beta_{012} \approx -\log 5$ and $\beta_{022} \approx 0$, and so,  (\ref{eq:DA}) wrongly classifies cluster 2 as DA. Let cluster $u=3$ be actually DA with $q^*_{i_13}=0.25$ and $q^*_{i_23}=0.05$ whenever $k_{i_1}=1$ and $k_{i_2}=2$. Then $\beta_{0k3} \approx -\log 2$ for $k=1,2$, and again, (\ref{eq:DA}) misclassifies   cluster 3 as non-DA. These  errors are avoided if non-DA cluster 2 is designated as the reference.

The following are some  practical strategies for choosing an appropriate reference cluster. We have used the second option in the simulation study and  data analyses. 
\begin{itemize}
    \item \textit{Minimum variance cluster} \quad 
    Adapting the technique of \cite{Nearing2022}, we fix the taxon for which the relative abundances are least variable  as the (singleton) reference cluster.
    
     \item \textit{Artificial reference taxon} \quad Augment  abundance matrix $\mathbf{Z}$ with an artificial  taxon with unit abundance counts for all $n$ samples. Since the taxa are exchangeable, label the  taxon as $j=1$ and  the remaining (actual) taxa as $j=2,\ldots,(p+1)$. Increment all the sampling depths, $L_1,\ldots,L_n$, by 1 to accommodate the additional count per matrix row. Assume that the artificial taxon constitutes its own cluster, and so  there are  $(H+1)$ clusters in the augmented dataset. Since the sampling depths of  microbiome datasets are many orders of magnitude greater than 1, this additional  ``taxon'' has small relative abundances, is therefore guaranteed to be non-DA, and   has virtually no  effect on the eventual inferences. 
\end{itemize}

Irrespective of the manner in which the reference cluster is chosen, the clusters are reordered so that  the reference  has the label $u=1$.

\subsection{Modeling non-biological zeros}\label{S:technical zeros}

Although the aforementioned aspects of the ZIBNP model are able to accommodate covariates, high-dimensionality, and compositionality, the challenge of high sparsity still remains. As discussed, biological zeros are detected  in a straightforward manner. Non-biological (i.e., sampling or technical) zeros require a more nuanced approach. Specifically, BNP submodel $\mathscr{F}$ accounts for 
sampling zeros, which  are therefore informative about the differential statuses.  By contrast,  technical zeros are caused by random sequencing  errors that  obfuscate important aspects of  $\mathscr{F}$.
We cast  technical zeros  in a missing data framework that  identifies the  source of non-biological sparsity to more accurately call the  differential taxa. 

Returning to the first mixture component in expression (\ref{eq:mixture}), 
we begin by modeling $r_{ij}$, the probability of technical zeros in sample $i$ and taxon $j$.  The log-sampling depths and  covariates are  known predictors of  technical zeros  \citep{technicalZero2020,Jiang2021}.   For a cluster-specific random effects vector, $\boldsymbol{\lambda}_{iu}=(\lambda_{0iu},\ldots,\lambda_{T+1,iu})'$,   
we therefore assume  
\begin{equation}
    \text{logit}(r_{ij}) = \lambda_{0ic_j} + \sum_{t=1}^{T} \lambda_{tic_j} x_{it} + \lambda_{T+1,ic_j} \log(L_i), \quad i=1,\ldots,n, \label{eq:zgp}
\end{equation}
where $\lambda_{tiu} \overset{\text{i.i.d.}}{\sim}  N(0,\tau_{\lambda}^2)$ for all  $t$, $i$, and $u$. Denote by $\boldsymbol{\Lambda}$ the collection of cluster-specific random matrices $\Lambda_1, \ldots, \Lambda_H$, where $\Lambda_u=((\lambda_{tiu}))$ is the $n \times (T+2)$ matrix of regression coefficients for the $u$th cluster    in equation (\ref{eq:zgp}). Because all $m_u$ taxa belonging to the $u$th cluster  have the same probability of technical zeros for a given sample $i$, we  write \[r_{ic_j}^* = r_{ij}, \quad j=1,\ldots,p.\]

From expressions (\ref{eq:mixture}) and (\ref{eq:sampling0}), and applying the CRP taxon-to-cluster mappings, the  probability of a non-biological (technical or sampling) zero for taxon $j$ of sample $i$   is then 
\begin{equation}
    P\bigl(Z_{ij}=0\mid c_j, \boldsymbol{\Lambda}, \mathbf{Q}^*\bigr) = r_{ic_j}^* + (1-r_{ic_j}^*) (1-q_{ic_j}^*)^{L_i}. \label{eq:Z=0}
    \end{equation}
  The lower-dimensional framework of submodel $\mathscr{F}$ and equation (\ref{eq:zgp}) imply that  the taxa in a   cluster  display somewhat similar zero patterns. 
  The next  innovation helps distinguish sampling zeros from technical zeros.

\paragraph{Censoring framework for  technical zeros}

We  interpret expression (\ref{eq:mixture})  as 
     an independent censoring mechanism that    converts the  true abundance count, $\Tilde{Z}_{ij}$, arising from submodel $\mathscr{F}$,  to an observed technical zero with probability $r_{ic_j}^*$. We denote the true, partially latent abundance matrix arising from  $\mathscr{F}$ by $\Tilde{\mathbf{Z}}=((\Tilde{Z}_{ij}))$. Parameters directly related to the DA taxa then rely on   matrix  $\Tilde{\mathbf{Z}}$ rather than    data matrix~$\mathbf{Z}$.
     Let $\delta_{ij}$ represent the  indicator that   $\Tilde{Z}_{ij}$ is  uncensored, so that $\delta_{ij} \stackrel{\text{indep}}\sim \text{Bernoulli}(1-r_{ic_j}^*)$.  The observed abundance counts are   related to the true counts as
\begin{equation}
    Z_{ij} = 
    \begin{cases}
        \Tilde{Z}_{ij} & \text{ if } \delta_{ij}=1 ,\notag\\
        0 & \text{ if } \delta_{ij}=0.\label{eqn:Z_delta}
    \end{cases} 
\end{equation}
If $Z_{ij}>0$, then clearly,  $\delta_{ij}=1$. On the other hand,
 $Z_{ij} = 0$   corresponds to either a sampling zero ($\delta_{ij}=1$) or technical zero ($\delta_{ij}=0$).

     Let latent set $J_i = \{j: \delta_{ij}=0, j=1,\ldots,p \}$ comprise  taxa with censored counts in the $i$th sample. Using latent matrix  $\Tilde{\mathbf{Z}}$, the observed sampling depths are  $L_i = \sum_{j \notin J_i} \Tilde{Z}_{ij}$. 
 The \textit{unobserved sampling depth} relies only on the censored taxa: $\tilde{S_i}= \sum_{j \in J_i} \Tilde{Z}_{ij}$. The \textit{true sampling depth} is  $\tilde{L}_i = \sum_{j =1}^p \Tilde{Z}_{ij}$ and  equals  $(L_i + \tilde{S_i})$.

 All aforementioned aspects  of the ZIBNP model, from equations (\ref{eq:mixture}) to (\ref{eq:Z=0}),  are  amended by replacing the observed  matrix $\mathbf{Z}$, count $Z_{ij}$, and sampling depth $L_i$, wherever they occur, with their  true (possibly latent) counterparts, i.e., true abundance matrix $\Tilde{\mathbf{Z}}$, true abundance counts $\tilde{Z}_{ij}$, and  true sampling depth   $\tilde{L}_i$.
Unlike traditional approaches for censored outcomes, however, censoring indicator $\delta_{ij}$ is unknown for the zero abundances, and the probability of a technical zero is \begin{equation}P\bigl[\delta_{ij}=0 \mid Z_{ij} = 0,  c_j, \tilde{L}_i, \boldsymbol{\Lambda}, \mathbf{Q}^*\bigr]=\frac{r_{ic_j}^*}{r_{ic_j}^* + (1-r_{ic_j}^*) (1-q_{ic_j}^*)^{\tilde{L}_i}}, \label{eq:technical?}
\end{equation}
with the conditional probability of a sampling zero given by the complementary event.
Since the sampling depths are usually  large,  a technical zero is  much more likely than a sampling zero, unless $q_{ic_j}^*$ is very small. In any case, equation (\ref{eq:technical?}) is applied to  call the  technical and sampling zeros (i.e., a posteriori generate the unknown censoring indicators) by the  MCMC~procedure outlined in Section \ref{S:MCMC}.

As mentioned, all inferences about parameters related to differential analysis rely on the partially latent, true abundance matrix $\Tilde{\mathbf{Z}}$. The following result, whose proof appears in Supplementary Material, facilitates posterior inferences about the true abundance matrix $\Tilde{\mathbf{Z}}$.

\begin{theorem} \label{prop_2_1}
Suppose  the censoring indicators $\delta_{ij}$ corresponding to the zero abundances  are  known. For sample $i=1,\ldots,n$,  let $\tilde{q}_i = \sum_{j \in J_i} q_{ic_j}^*$. Then
 \begin{enumerate}
     \item \label{prop2_1_1} True sampling depth $\tilde{L}_i$ has a negative binomial distribution:
     \[\tilde{L}_i \mid J_i, \mathbf{c}, \mathbf{Q}^* \sim \text{NegBin}\bigl(L_i,\, 1-\tilde{q}_i\bigr),
     \]
     for which $\tilde{L}_i$ is the random number of i.i.d.\ Bernoulli trials with success probability  $(1-\tilde{q}_i)$ and     
      $L_i$ is the prespecified number of successes. Hence, $\tilde{S}_i=\tilde{L}_i-L_i$.
     \item \label{prop2_1_2} Let the vector of $|J_i|$ unobserved taxa abundances be  $\Tilde{\mathbf{Z}}_i^{(0)}=\bigl(\Tilde{Z}_{ij}:\delta_{ij}=0, \, j=1,\ldots,p\bigr)$. Define $w_{ij}=q_{ic_j}^*/\tilde{q}_i$ for $j \in J_i$, and probability vector $\boldsymbol{w}_i= (w_{ij}: j \in J_i)$ of length $|J_i|$. Then 
    \[\Tilde{\mathbf{Z}}_i^{(0)} \mid \tilde{L_i}, J_i, \mathbf{c}, \mathbf{Q}^* \sim \text{Multinomial}\bigl(\tilde{S}_i, \,\boldsymbol{w}_i\bigr),
      \]
      where $\tilde{S}_i=\tilde{L}_i-L_i$.
       \end{enumerate}
    
\end{theorem}

\begin{figure}
    \centering
    \includegraphics[width=\textwidth]{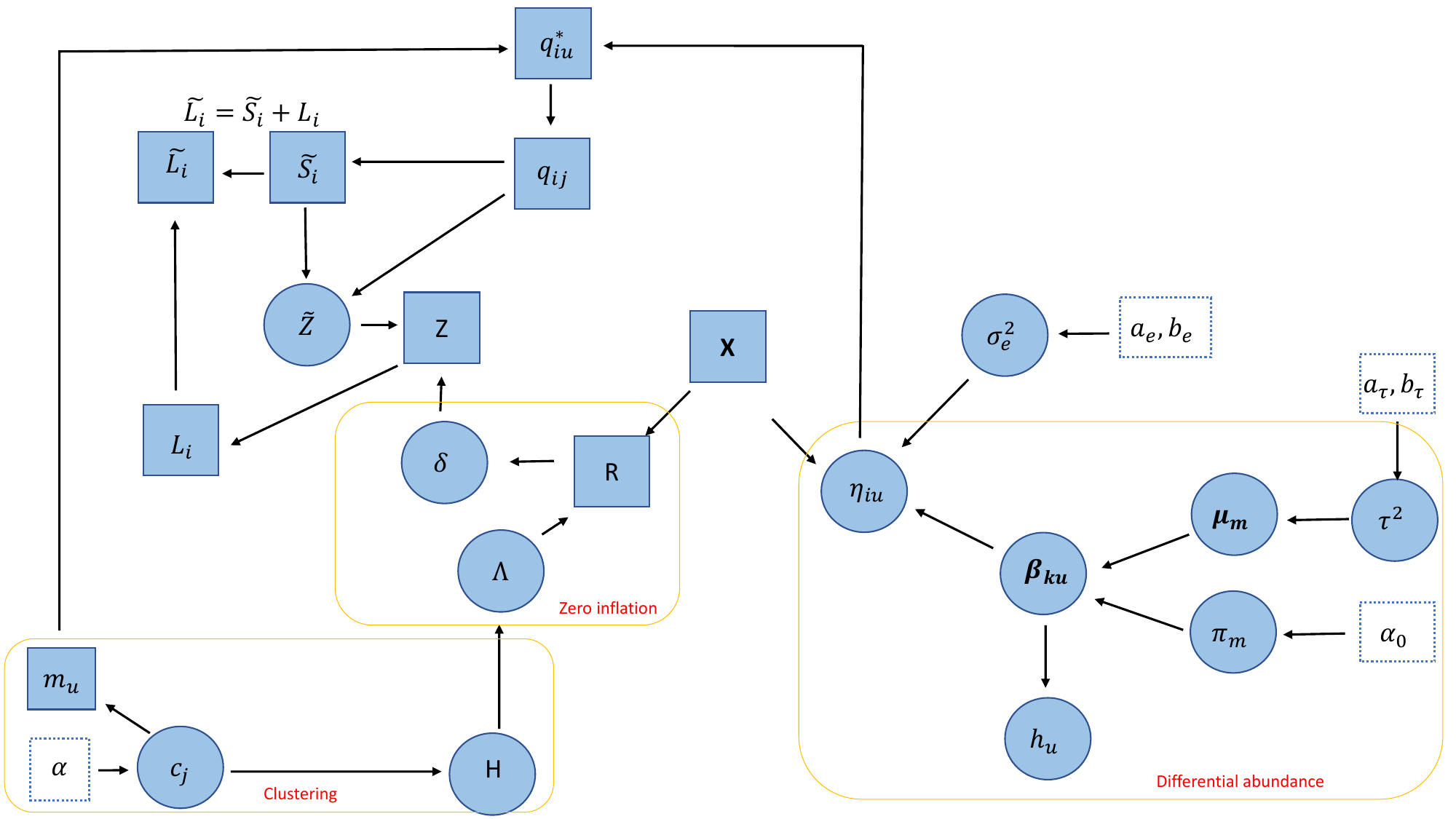}
    \caption{DAG representation of the ZIBNP model highlighting  parameters related to handling sparsity (``zero inflation''), high-dimensionality (``clustering''), and DA taxa detection (``differential abundance''). 
Circles represent stochastic  parameters, solid rectangles represent the data and deterministic variables, and open rectangles represent prespecified constants. }
    \label{fig:dag_zibnp}
\end{figure}

Finally, we complete the ZIBNP model by assigning  standard conjugate priors to the remaining  hyperparameters. Figure \ref{fig:dag_zibnp} presents a directed acyclic graph (DAG)  of the parameters.

\section{Posterior inferences}\label{S:inference}

Section \ref{S:MCMC} outlines the Monte Carlo inference procedure. Section \ref{S:posterior DA taxa} describes the technique for detecting the DA taxa.

\subsection{MCMC  algorithm}\label{S:MCMC}
After initialization by  naive estimation techniques, as described in  Supplementary Material, the model parameters are  iteratively updated using  MCMC procedures. All   parameters except the $\eta_{iu}$'s are updated by Gibbs sampling. Although the full conditional  of $\eta_{iu}$ does not have a closed form,  it is log-concave, and  $\eta_{iu}$ can therefore be generated by adaptive rejection sampling \citep{GilksWild1992}.  See Supplementary Material for an outline of the MCMC steps. The post-burn-in MCMC sample of the parameters  (including mapping variables, group-cluster regression coefficients, censoring indicators,  true abundance counts with imputed technical zeros, and differential statuses) is stored and processed for posterior inferences.

\subsection{Calling taxa differential statuses}\label{S:posterior DA taxa}

Using $L$ post-burn-in MCMC samples,
a straightforward estimator of $P\bigl[\tilde{h}_j=1 \mid \mathbf{Z}\bigr]$,   the posterior probability that taxon $j$
is non-DA, is  $\sum_{l=1}^L \mathcal{I}(\tilde{h}_j^{(l)}=1)/L$.  However, the following theoretical result gives a more precise ``Rao-Blackwellized''   estimator that is  especially accurate for taxa with ambiguous differential statuses. But first, for group $k$ and non-reference latent cluster $u>1$, let \textit{membership variable} $v_{ku}$  document the $M$-mixture component  (\ref{eq:beta}) from which   regression vector $\boldsymbol{\beta}_{ku}$ is drawn. That is, \begin{equation}
    \boldsymbol{\beta}_{ku}= \boldsymbol{\mu}_{v_{ku}}, \quad k=1,\ldots,K,\, u=2,\ldots,H.\label{eq:v_ku}
\end{equation}
The theorem follows from the law of total probability. We omit the proof for brevity. 

\begin{theorem}\label{thm 2}
Let the set $\Theta^-$  contain  all ZIBNP model parameters except the mapping variables, differential status variables for the  taxa and clusters, membership variables (\ref{eq:v_ku}), and any  parameters having a deterministic relationship with these variables. Let   $P_*[\cdot]$ denote the  conditional  probability, $P\bigl[\cdot \mid \Theta^-, H, 
 \mathbf{Z}\bigr]$.  Then
\begin{enumerate}[itemsep=5pt,parsep=5pt]
    \item\label{thm 2:part 1} For  non-reference cluster $u$, the  conditional  posterior probability that a  cluster is non-DA is \[P_*\bigl[h_u=1\bigr]= \sum_{m=1}^{M} \prod_{k=1}^{K} P_*\bigl[v_{ku}=m\bigr], \quad u=2,\ldots,H.
    \]
   
    \item\label{thm 2:part 2} For  $j=1,\ldots,p$, the  conditional posterior  probability that the $j$th taxon is non-DA is \[P_*\bigl[\tilde{h}_j=1\bigr]= P_*[c_j=1] + \sum_{c=2}^{H}   P_*[h_u=1] P_*[c_j=u].
    \]
\end{enumerate}
\end{theorem}
Applying Theorem~\ref{thm 2}, an estimator for the  posterior probability that the $j$th taxon is non-DA is then
\[
\widehat{P}\bigl[\tilde{h}_j=1 \mid \mathbf{Z}\bigr]= \frac{1}{L}\sum_{l=1}^{L}  P\bigl[\tilde{h}_j=1 \mid \Theta_{(l)}^-, H_{(l)}, 
 \mathbf{Z}\bigr],
\]
where $\Theta_{(l)}^-$ and $H_{(l)}$ are the parameter values generated at the $l$th MCMC iteration. Apply the Rao-Blackwell theorem, it is easy to verify that this estimator is more precise than $\sum_{l=1}^L \mathcal{I}(\tilde{h}_j^{(l)}=1)/L$ even though both estimators are consistent as $L \to \infty$.

The procedure is applied to infer the posterior probabilities of the taxa differential statuses  along with  uncertainty estimates. Given a nominal FDR (say, 5\%),
we apply the Bayesian FDR  approach  of \cite{pmid15054023} to choose  an appropriate  threshold, $\kappa$, for calling the DA taxa.  
 An estimate of  differential status variable $\tilde{h}_j$ is  
\[
\widehat{\tilde{h}}_j = 1+ \mathcal{I}\biggl(\widehat{P}\bigl[\tilde{h}_j=1 \mid \mathbf{Z}\bigr]\ge \kappa\biggr), \quad j=1,\ldots,p,
\]
where $\mathcal{I}(\cdot)$ represents the indicator function. This gives the estimated set of DA taxa,  $\hat{\mathscr{D}}=\bigl\{j: \widehat{\tilde{h}}_{j}=2,  \,\, j=1,\ldots,p\bigr\}$.

\section{Simulation studies}\label{S:simulation}
We investigated the accuracy of the proposed ZIBNP approach  using artificial microbiome datasets with different sparsity levels. For comparison, we analyzed the data using existing statistical methods  with available R packages capable of detecting DA taxa in the presence of covariates.  

\paragraph*{Data generation} \quad The following steps, which differ in several important aspects from the ZIBNP model,  were used to generate the artificial datasets. For $ n=100$ subjects belonging to $K=2$ groups,  we generated   abundance matrix $\mathbf{Z}$ for $p=1,000$ taxa, with   taxon 1 constituting an artificial reference  cluster  with unit abundance counts.  
 For four sparsity levels modulated by a simulation parameter~$\lambda_0$ in Step \ref{lambda_0} below, we independently generated $25$ datasets as follows: 
    
\begin{enumerate}[itemsep=5pt,parsep=5pt,label=(\alph*)]
    \item \textbf{Covariates} \quad We used   the actual  covariates  of a publicly available  oral microbiome dataset \citep{Burcham2020} consisting of adults and children as the subject groups. We  selected $T=4$ covariates, namely, three binary covariates (indicators of antibiotics intake during the last 6 months, adequate  brushing habits, and sex)  and one continuous  covariate (BMI). 
    Randomly sampling the covariates of $50$ subjects,  we replicated the \textit{same} covariates in adults and children to obtain the  covariate matrix, $\mathbf{X}$, of dimension $100$ by $4$, of the artificial dataset. That is, if  $\mathbf{X}^{(0)}$ denotes the $50 \times 4$ covariate matrix randomly  sampled from the  oral microbiome dataset, then 
    \begin{equation}
        \mathbf{X} = \left[
    \begin{matrix}
    \mathbf{X}^{(0)}\\
    \mathbf{X}^{(0)}\\
    \end{matrix}
    \right], \label{eq:sim_X}
    \end{equation}
    where the first 50 subjects represent adults and the remaining 50 subjects represent children in the artificial dataset.
    Although unnecessary for the downstream analysis, this structure helps informally validate the detected DA taxa, as   discussed later. 
    
    \item \textbf{Mapping variables}\quad We generated the true number of clusters, $H$, from a discrete uniform distribution on  $\{8,9,\ldots,20\}$. Applying the \texttt{rpartitions} package in R, we generated a $H$-partition of $p$ objects using the  \texttt{rand\_partitions} function of the package. The true taxon-to-cluster mapping variables, $c_1,\ldots,c_p$, were set equal to this randomly generated partition and therefore relied on a very different stochastic mechanism than  CRP (\ref{eq:CRP}).

    \item \textbf{True differential statuses}\label{simulation_h}\quad  In relation (\ref{eq:beta}), setting $M=7$ and $\tau^2=1$, and since $(T+1)=5$, the finite mixture locations, $\boldsymbol{\mu}_m$, were independently generated from  distribution $N_{5}\bigl(\mathbf{0}, ({\mathbf{X}^{\dag}}^T\mathbf{X}^{\dag})^{-1}\bigr)$, for $m=1,\ldots,7$. The mixture probability vector $\boldsymbol{\pi}$ was generated from the Dirichlet distribution, $\text{Dir}_{7}\bigl(\frac{1}{7}\mathbf{1}_7\bigr)$. For the $(H-1)$ non-reference clusters, the $K(H-1)=$ $2(H-1)$ regression vectors, $\boldsymbol{\beta}_{ku}$, were generated from  finite mixture model  (\ref{eq:beta}) consisting of $M=7$ components. The true differential statuses $h_2,\ldots,h_H$ of the non-reference clusters were calculated by applying (\ref{eq:DA}). As usual, $h_1=1$ for the reference cluster. The true set of DA taxa, $\mathscr{D}$, was evaluated using (\ref{eq:D}).

    \item \textbf{Probability matrix $\mathbf{Q}$} \quad Unlike  the ZIBNP model, in order to generate the motif elements of the artificial datasets, we assumed 
    \begin{equation}
        q_{iu}^* = \alpha_{ih_u} \exp(\zeta_{iu}), \quad u>1, \label{eq:qstar}
    \end{equation}
    where $\alpha_{ih_u}$ is the subject-specific proportionality constant that depends on the DA status of cluster $u$, and the   $\zeta_{iu}$'s  were generated as
    \begin{equation}
 \zeta_{iu}  \stackrel{\text{indep}}\sim N\biggl(\beta_{0k_iu} + \sum_{l=1}^Tx_{il} \beta_{lk_iu}, \,\, \sigma_{\zeta}^2\biggr),   \quad \text{$i=1,\ldots,n$, and $u>1$,}  \label{eq:zeta}
  \end{equation}
  with $\sigma^2_{\zeta}$ chosen so  that  $R^2$ exceeded $99\%$. 
    Since  microbiome data are compositional, $\sum_{i=1}^{p} q_{ij} = \sum_{u=1}^{H} m_u q_{iu}^* = 1$, so that  
    \begin{equation} \sum_{u : h_u =1} \alpha_{i1} m_u \exp(\zeta_{iu}) + \sum_{u : h_u =2} \alpha_{i2} m_u \exp(\zeta_{iu})\label{eq:qstar2} =1.\end{equation}
    For $h=1,2$, let $\rho_{ih}=\sum_{u : h_u =h} \alpha_{ih} m_u \exp(\zeta_{iu})$ be the total probability assigned to the non-DA and DA taxa, respectively, in equation (\ref{eq:qstar2}).  
     We generated $\rho_{i1}  \sim \text{beta}(0.5A,0.5A)$, for $A$ large, and set $\rho_{i2}=1-\rho_{i1}$. Thereby, we computed
    \begin{equation}
        \alpha_{ih} = \frac{\rho_{ih}}{\sum_{u : h_u =h} m_u \exp(\zeta_{iu})}, \quad h=1,2,  \label{eq:prop}
    \end{equation}
    to evaluate the matrix $\mathbf{Q}^*$.
      Then, applying equation (\ref{eq:qtilde}), we obtained probability matrix~$\mathbf{Q}$.
      
    \item \textbf{Uncensored taxa abundances}\quad To mimic the range of sampling depths observed in actual microbiome datasets, we  generated the true sampling depths, $\tilde{L}_i \stackrel{\text{i.i.d.}}\sim \text{Poisson}(10,000)\times \text{Poisson(100)} $, for subjects $i=1,\ldots,100$.
    The true taxa abundances were  generated as $\boldsymbol{\tilde{Z}}_i \sim \text{Multinomial}(\tilde{L}_i, \mathbf{q}_i)$.
    
    \item\label{lambda_0} \textbf{Observed taxa abundances}\quad Setting $\boldsymbol{\lambda}_{iu}= \lambda_0 \boldsymbol{1}$, where $\lambda_0 \in \mathcal{R}$, and using the true sampling depths $\tilde{L}_i$, we applied equation (\ref{eq:zgp})  to compute the probability of technical zeros, $r_{iu}^*$, for $u=2,\ldots,H$. We  generate the  censoring indicators   as $\delta_{ij} \stackrel{\text{indep}}\sim \text{Bernoulli}(1-r_{ic_j}^*)$ for $j=1,\ldots,p$. Thereafter, we  calculated the taxa abundance matrix $\mathbf{Z}$ using (\ref{eqn:Z_delta}). Varying simulation parameter $\lambda_0$ in the set $\{-0.1, -.059, -.001, .023\}$ produced a range of sparsity levels  matching typical microbiome data; see Table \ref{tab:zeroperc}. 
\end{enumerate}

\begin{table}[]
    \centering
    \renewcommand{\arraystretch}{1.2}
    \begin{tabular}{@{}ll@{}}
    \hline
    $\lambda_0$ & \% zeros \\
    \hline
        -0.100 & 13\% \\
         -0.059 & 25\% \\
         -0.001 & 50\% \\
         0.023 & 60\% \\
         \hline
    \end{tabular}
    \caption{Averaging over the 25 artificial datasets, observed sparsity in   abundance matrix $\mathbf{Z}$ for  different $\lambda_0$ values.}  \label{tab:zeroperc}
\end{table}

\begin{figure}
    \centering
    \includegraphics[scale=0.45]{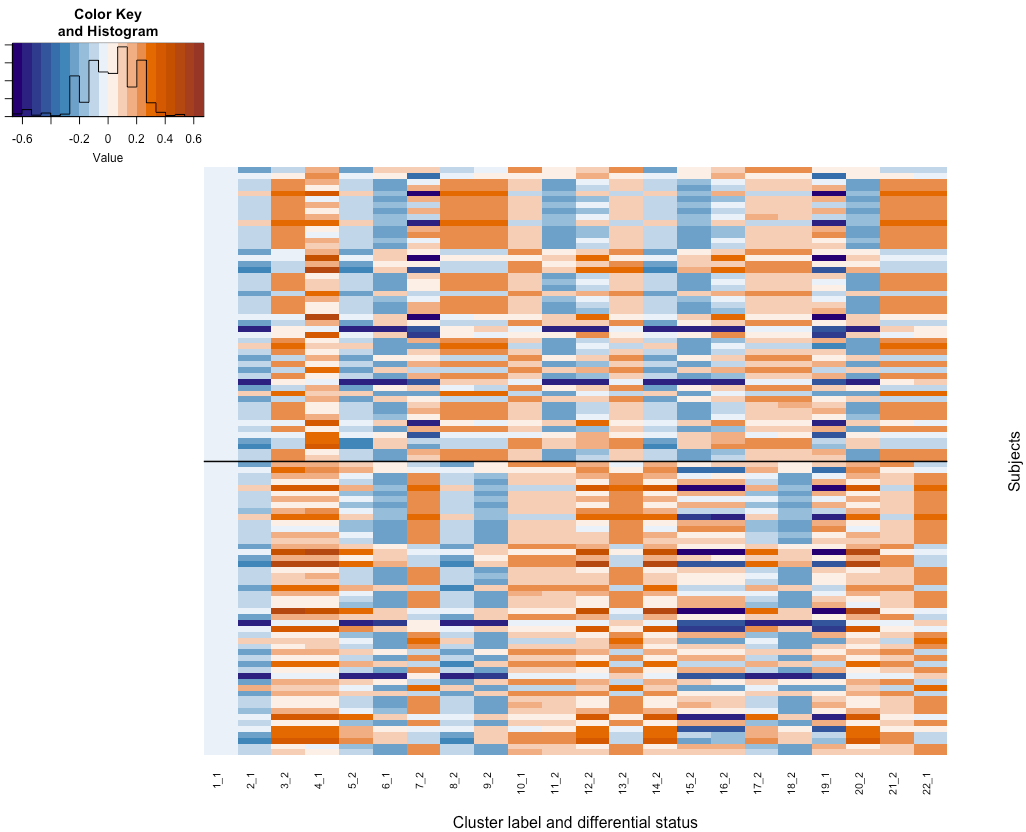}
    \caption{For a randomly chosen artificial dataset  with $H=22$ clusters, heat map of the true log ratios, $\zeta_{iu}$, for subject $i$ (rows) and cluster $u$ (columns), underlying the generated taxa abundances. 
    The first 50 subjects are adults (group 1) and the remaining 50 subjects are children (group 2). The horizontal line separates the two groups of subjects. The column labels in the lower axis have the format ``$u\_{h_u}$,'' for $u=1,\ldots,H$, to indicate  the  index and  true differential status of the $u$th cluster. Reference cluster~1 has  column label ``$1\_1$'' and is  non-DA by design.  See Section \ref{S:simulation} for further discussion.}
    \label{fig:heatmap_eta}
\end{figure}

Figure \ref{fig:heatmap_eta} displays a heat map of the true log ratios, $\zeta_{iu}$, for the $H=22$ true clusters of a randomly chosen dataset. The $p=1,000$ taxa are not shown in the figure.  
Due to special structure (\ref{eq:sim_X}),  motif elements (\ref{eq:qstar}), and relatively small $\sigma^2_{\zeta}$ in (\ref{eq:zeta}), we can accurately call the true differential statuses of the non-reference clusters by a visual comparison of  the color patterns of  the two groups in Figure \ref{fig:heatmap_eta}. Specifically, for $u=1,\ldots,H$, the $u$th cluster, represented by the $u$th column in Figure \ref{fig:heatmap_eta}, is non-DA if the color patterns in the upper and lower column halves are identical, in which case the column label is ``$u\_1$''; otherwise, the cluster is DA and its column label is ``$u\_2$''. For example, clusters $1$ and $2$ are non-DA ($h_1=h_2=1$) because the patterns are identical for adults and children. Cluster $3$ is DA ($h_3=2$).

Disregarding  knowledge of the  generation mechanism, the Section~\ref{S:inference} procedure was applied to analyze each artificial dataset and make posterior inferences   using the ZIBNP model. Applying the Section \ref{S:posterior DA taxa} strategy, we post-processed the MCMC sample to 
 estimate the DA posterior probabilities, $P\bigl[\tilde{h}_j=2\mid \mathbf{Z}\bigr]$. Varying the posterior probability thresholds for calling the DA taxa, we compared $\hat{\mathscr{D}}$, the detected DA taxa using that  threshold, with the true $\mathscr{D}$; see Step \ref{simulation_h}. The sensitivities and specificities over the  range of thresholds were evaluated to produce the ROC curve for each dataset. Additionally, as described in Section \ref{S:posterior DA taxa}, assuming a nominal FDR of 5\%,
we applied the Bayesian FDR procedure to call the DA taxa, and thereby, evaluated the  FDR and sensitivity achieved by ZIBNP for each simulation  dataset.

Using area under the curve (AUC), false discovery rate (FDR), and sensitivity as the evaluation criteria, we  compared our technique with  some well-established methods for differential abundance analysis  with covariates, focusing only on   methods implemented in publicly available R packages:
 \begin{enumerate}[label=(\roman*)]
     \item ANCOMBC \citep[version 1.0.5;][]{Lin2020}: We used the \texttt{ancombc} function  with default settings. The function fits  compositional  data by making a log transformation and using a sample-specific offset term for bias correction.
     
     \item Maaslin2 \citep[version  1.4.0;][]{maaslin2_2021}: The \texttt{Maaslin2} function was used with total sum scaling normalization and covariate fixed effects.  
     This technique fits a generalized linear model to each feature abundance with respect to the covariates, checking for significance using the Wald test.
     
     \item Metagenomeseq  \citep[version 1.32.0;][]{Paulson2013}: 
Function \texttt{fitZIG}  was applied to fit a zero-inflated Gaussian version of the  technique. The workflow involved normalizing the data using cumulative sum scaling  with  default settings.   

\item DESeq2 \citep[version 1.30.1;][]{Love2014}:
We used the \texttt{DESeq2} function  with  argument {\tt sizefactor} set to ``poscounts.'' The R function performs a likelihood ratio test for a reduced model containing all covariates except the grouping factor. 
 \end{enumerate}

\begin{figure}[h!]
    \centering
    \includegraphics[width=\textwidth]{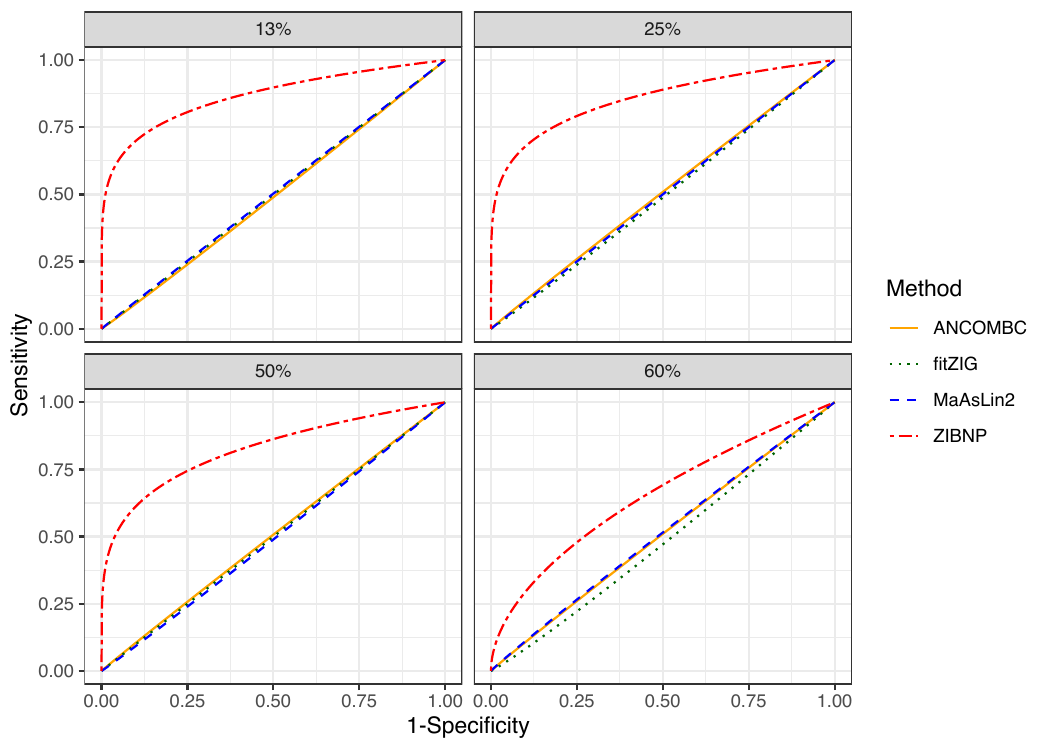}
    \caption{For the simulation study, ROC plots comparing the methods ZIBNP, ANCOMBC, fitZIG, and MaAsLin2. The  panels represent different sparsity levels ranging from 13\% to 60\% zeros.}
    \label{fig:sim_roc_all}
\end{figure}

The FDRs of the competing statistical methods were  controlled via   taxa-specific q-values, i.e., adjusted p-values \citep{BenjaminiHochberg_1995}.
 Using the q-values, a method's sensitivities and specificities for critical values between 0 to 1  produced the ROC curve for each dataset. Additionally, for a target FDR of 5\%, the DA taxa were called using the  Benjamini-Hochberg procedure   to evaluate a method's achieved FDR and sensitivity for each  dataset.

 Averaging over the 25 datasets, Figure \ref{fig:sim_roc_all}  displays the   ROC curves of all the  methods, with the panels corresponding to the four sparsity levels in Step \ref{lambda_0}.   The corresponding average AUCs, along with 95\% confidence intervals for AUC, are presented  in Table~\ref{tab:auc_table_sim}. Even though the accuracy of ZIBNP  deteriorated with increasing sparsity as expected,  we  find that ZIBNP has a substantially higher AUCs than the competing methods for all sparsity levels.

\begin{figure}
    \centering
    \includegraphics[scale=0.6]{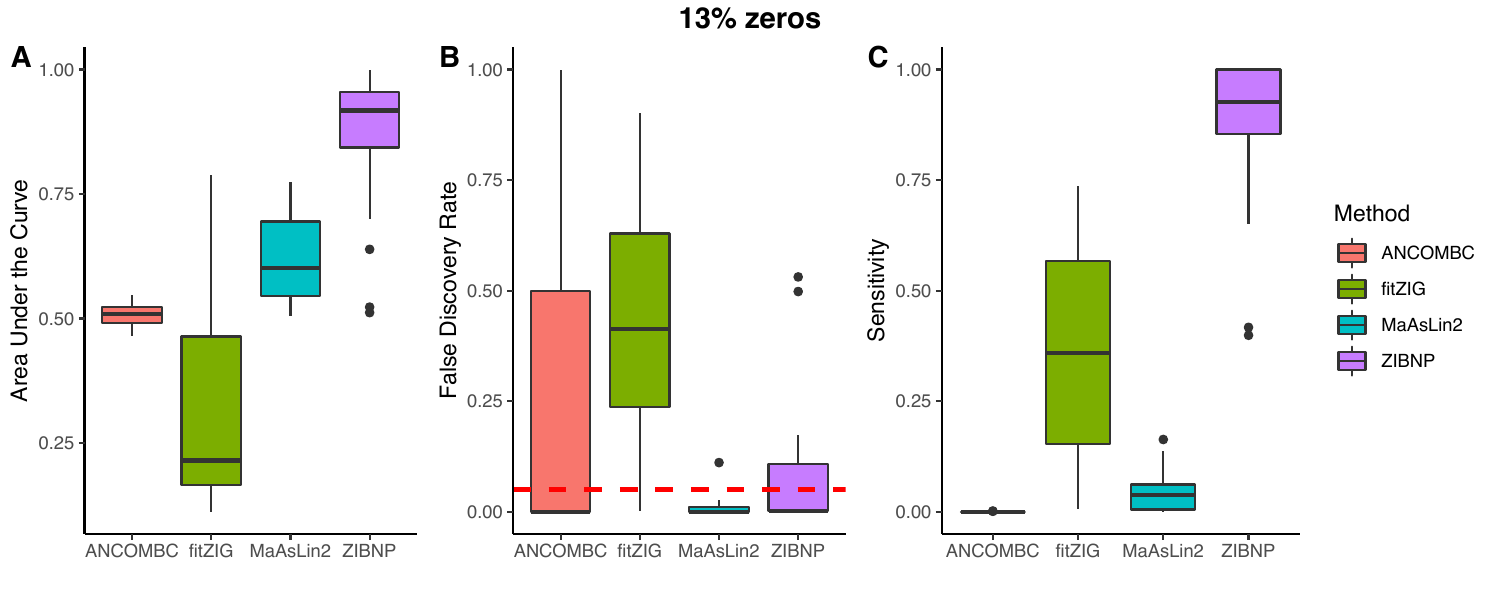}
    \caption{For 13\% zeros in the abundance matrix, summarizing over the 25 artificial datasets of Section \ref{S:simulation}, boxplots of AUC (A), FDR (B), and sensitivity (C) for the different statistical methods. The dashed horizontal line in sub-figure~B indicates the  target FDR of 5\% for all the methods.}
    \label{fig:perf_case1}
\end{figure}

\begin{figure}
    \centering
    \includegraphics[scale=0.6]{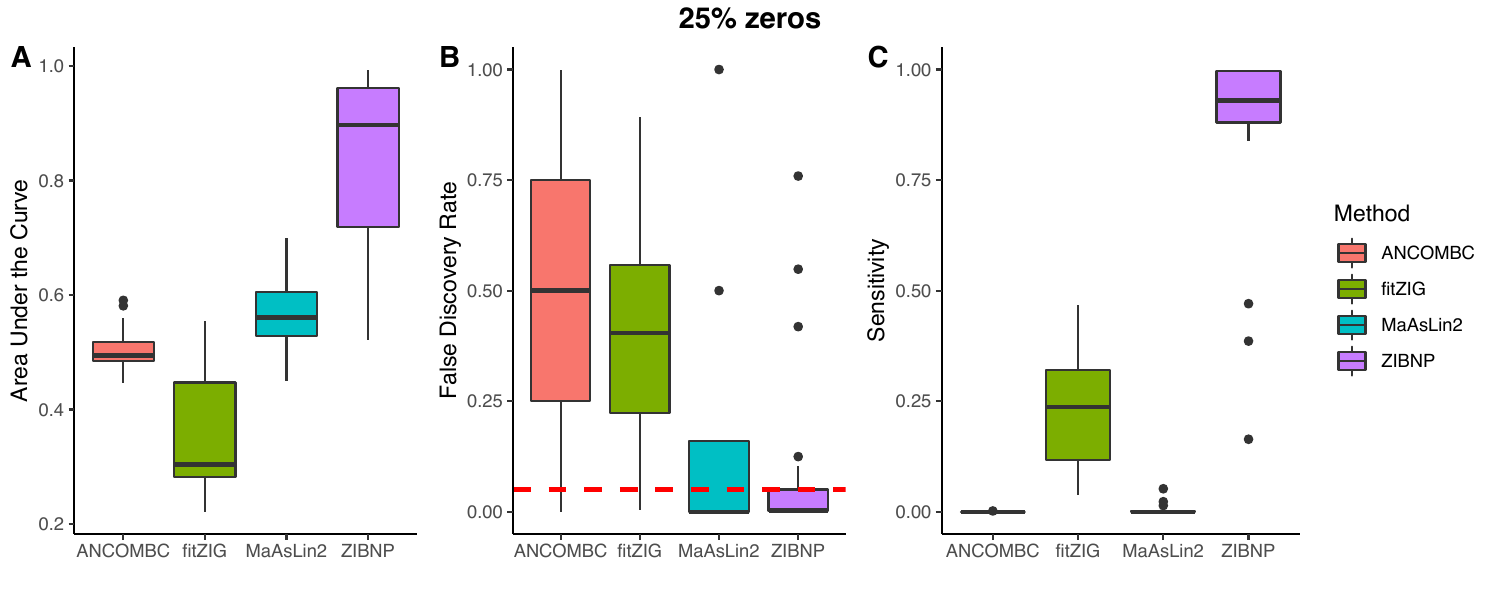}
    \caption{For 25\% zeros in the abundance matrix, summarizing over the 25 artificial datasets of Section \ref{S:simulation}, boxplots of AUC (A), FDR (B), and sensitivity (C) for the different statistical methods. The dashed horizontal line in sub-figure~B indicates the  target FDR of 5\%  for all the methods. }
    \label{fig:perf_case2}
\end{figure}

 For  $13\%$ and $25\%$ sparsity, Figures \ref{fig:perf_case1} and \ref{fig:perf_case2} respectively represent the AUC, FDR, and sensitivity for the statistical methods.  The method DESeq2   declared all  taxa as non-DA in the simulated datasets, and is not shown in the plots. 
In both plots, 
  ANCOMBC had a median AUC  of around  $0.5$ (sub-figure A), exhibited relatively high FDR (sub-figure B) and low sensitivity (sub-figure C). 
  For fitZIG, 
  the typical AUC in sub-figure A  was  below $0.5$  
  with an elevated FDR in sub-figure B, but its sensitivity was relatively high (sub-figure C). The AUCs for MaAsLin2 were relatively high and the  FDRs were well below the target FDR, but  the sensitivity was somewhat low. Further investigation into the performance of MaAsLin2 revealed that for most of the 25 artificial datasets and all sparsity levels, the technique failed to  detect most of the DA taxa. These results are consistent with previous studies that  found similar AUCs and FDRs for fitZIG and MaAsLin2 \citep[e.g.,][]{Thorsen2016,Hawinkel2017, maaslin2_2021}.

By contrast, the figures and tables reveal that ZIBNP displayed substantially higher AUC, well-controlled FDR, and high sensitivity. 
In summary, ZIBNP  appreciably outperformed the methods DESeq2, ANCOMBC, fitZIG, and MaAsLin2 in  inferential accuracy and  was reliable even with  highly sparse  datasets.

\section{Data analysis}\label{sec:dataAnalysis}

We applied the  ZIBNP technique to analyze  publicly available microbiome data from the CAMP  and Global Gut  studies, consisting of $K=2$ and $K=3$ groups, respectively.  The results were compared with other differential analysis techniques.

\begin{table}[]
    \centering
    \begin{tabular}{@{}lllll@{}}
    \hline
         &  \multicolumn{4}{c}{\textbf{Percentage of Zeros}} \\
         \hline
          & \textbf{13\%} & \textbf{25\%} & \textbf{50\%} & \textbf{60\%}  \\
         \hline
        ZIBNP & {0.79} (0.67,0.95) & {0.77} (0.67,0.96) & {0.68} (0.55,0.83) & 0.58 (0.44,0.69) \\
        ANCOMBC & 0.49 (0.43,0.54) & 0.51 (0.45,0.62) & 0.51 (0.45,0.58) & 0.51 (0.47,0.64) \\
        fitZIG & 0.50 (0.41,0.62) & 0.49 (0.40,0.56) & 0.50 (0.39,0.61) & 0.50 (0.45,0.54)  \\
        MaAsLin2 & 0.50 (0.41,0.59) & 0.50 (0.46,0.58) & 0.50 (0.43,0.56) & 0.51 (0.47,0.60) \\
        \hline
    \end{tabular}
    \caption{For the simulation study, aggregating over the 25 artificial datasets, average  and 95\% confidence intervals of the ROC plot AUCs for different DA  methods (rows). The columns correspond to  different data sparsity levels.}
    \label{tab:auc_table_sim}
\end{table}

\paragraph*{CAMP study canine data} \quad
The dataset was downloaded from the MicrobiomeDB resource \citep{mdb}. To investigate the association between eukaryotic parasite infection and the gut microbiome, the investigators processed fecal samples of 155 infected (case) and 115 uninfected (control) dogs and sequenced the V4 region of the 16S rRNA gene.  Animal-specific attributes such as sterilization, pet ownership, age, and sex were recorded, allowing statistical methods  to  adjust for  these covariates while detecting the DA taxa.

Table \ref{tab:CAMP_dataxa} presents the DA taxa  detected by ZIBNP between the case and control groups. Some of the detected taxa have been reported by previous studies. For example, bacteria of the genus \textit{Bacteroides} were detected as DA with low relative abundance in infected dogs. In studies of the effect of parasite-induced infections on gut microbiome composition,  bacteria of the genera \textit{Megamonas} and \textit{Prevotella} were significantly associated  with the infection status of dogs  \citep{Berry2020,dogs_prevotella}. The genus \textit{Blautia} has been  reported to have a strong dysbiosis in dogs with acute diarrhea  \citep{dogs_prevotella}. In addition, microbes belonging to the genera \textit{Streptococcus} \citep{dogs_prevotella}, \textit{Ruminococcus gnavus} \citep{Ruminococcus_ibd,dog_giardia}, and \textit{Alloprevotella} \citep{dog_giardia} have known associations with  inflammatory bowel disease.

\begin{figure}
    \centering\includegraphics[width=0.6\textwidth]{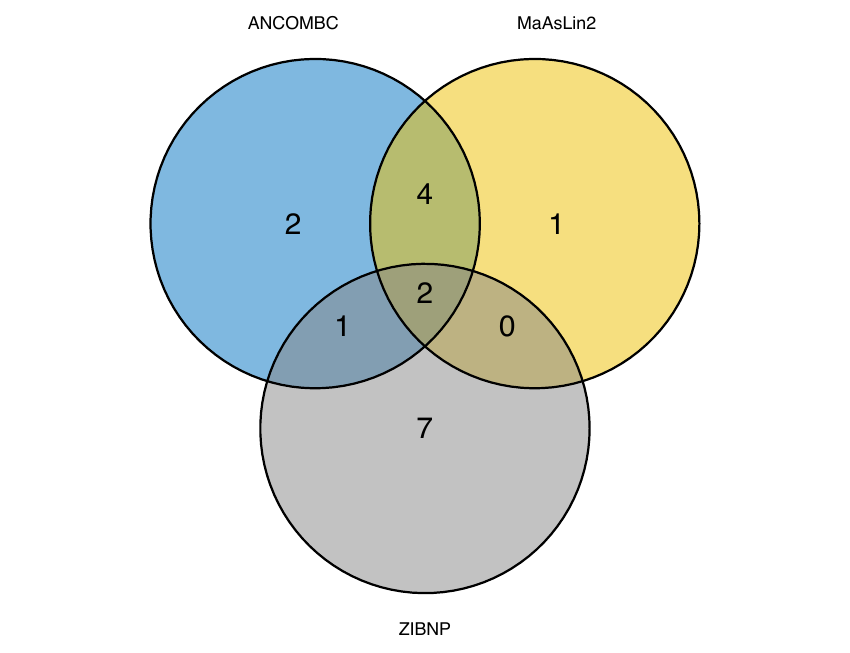}
    \caption{Venn diagram of the number of DA taxa detected by the methods ANCOMBC, MaAsLin2, and ZIBNP in the CAMP canine dataset. See the text for further discussion.}
    \label{fig:my_label}
\end{figure}

The methods ANCOMBC, fitZIG, Maaslin2, DESeq2, and ZIBNP detected 9, 72, 7, 42, and 10 DA taxa, respectively. Table \ref{tab:jaccard_camp} displays the pairwise similarity measures between the different sets of DA taxa detected by the  statistical methods using the  Jaccard index \citep{Jaccard1901}. The Jaccard index ($J$) is a measure of similarity  between two sets that ranges between $0\%$ to $100\%$, and is defined as the size of the intersection set divided by the size of the union set. Whereas $J = 0\%$ implies that  the two sets have no overlap,  $J=100\%$ implies the two sets are identical. Although  all the methods in Table~\ref{tab:jaccard_camp} display low overlap, there is moderate overlap between ANCOMBC and Maaslin2, possibly because both  methods rely on generalized linear models. The proposed ZIBNP method shares some DA taxa with ANCOMBC and Maaslin2. However,  ZIBNP has a low overlap with  fitZIG and DESeq2, which respectively rely on zero-inflated Gaussian and  negative binomial models. A Venn diagram of the number of the DA taxa detected by ANCOMBC, MaAsLin2, and ZIBNP is presented in Figure~\ref{fig:my_label}. The genera \textit{Blautia} and \textit{Lachnoclostridium} were the common findings of ZIBNP, ANCOMBC, and Maaslin2. Additionally,  \textit{Ruminococcus gnavus}  was  detected by  both  ZIBNP and ANCOMBC.

\begin{sidewaystable}
    \centering
    \resizebox{\textwidth}{!}
    {\begin{tabular}{lllllll}
\hline
Kingdom &	Phylum &	Class &	Order &	Family &	Genus &	Species \\
\hline
Bacteria &	Bacteroidota &	Bacteroidia &	Bacteroidales &	Bacteroidaceae &	Bacteroides &	- \\
Bacteria &	Bacteroidota &	Bacteroidia &	Bacteroidales &	Prevotellaceae &	Alloprevotella &	 -\\
Bacteria &	Bacteroidota &	Bacteroidia &	Bacteroidales &	Prevotellaceae &	Prevotella &	 -\\
Bacteria &	Firmicutes &	Bacilli &	Lactobacillales &	Streptococcaceae &	Streptococcus &	 - \\
Bacteria &	Firmicutes &	Clostridia &	Clostridiales &	Clostridiaceae & Clostridium sensu stricto 1	 &	-  \\
Bacteria &	Firmicutes &	Clostridia &	Lachnospirales &	Lachnospiraceae &	[Ruminococcus] gnavus group & -	 \\
Bacteria &	Firmicutes &	Clostridia &	Lachnospirales &	Lachnospiraceae &	Blautia &	 -\\
Bacteria &	Firmicutes &	Clostridia &	Lachnospirales &	Lachnospiraceae &	Lachnoclostridium &	- \\
Bacteria &	Firmicutes &	Clostridia &	Peptostreptococcales-Tissierellales &	Peptostreptococcaceae & Peptoclostridium &	- \\
Bacteria &	Firmicutes &	Negativicutes &	Veillonellales-Selenomonadales &	Selenomonadaceae &	Megamonas &	Megamonas funiformis \\
\hline
        \end{tabular}}
        \caption{In the CAMP study,  differentially abundant taxa detected by ZIBNP between infected and uninfected dogs as the groups.}
    \label{tab:CAMP_dataxa}
 \end{sidewaystable}

\vspace{2em}

\begin{table}[]
\centering
\begin{tabular}{llllll}
\hline
& ANCOMBC  & fitZIG   &    Maaslin2 &  DESeq2  &  ZIBNP  \\
\hline
ANCOMBC  &1 & 0.03 & 0.6 & 0.11 & 0.19 \\
fitZIG  & 0.03 & 1 & 0.03 & 0.28 & 0 \\   
Maaslin2  & 0.6 & 0.03 & 1 & 0.11 & 0.13 \\
DESeq2  & 0.11 & 0.28 & 0.11 & 1 & 0.02 \\
ZIBNP  & 0.19 & 0 & 0.13 & 0.02 & 1  \\
\hline
\end{tabular}
\caption{For the CAMP study data, pairwise Jaccard index of the DA taxa  detected by different statistical methods.}
\label{tab:my_label} \label{tab:jaccard_camp}
\end{table}

\subsection{Global Gut microbiome study}

We applied the proposed ZIBNP method to analyze the motivating Global Gut microbiome data \citep{Yatsunenko2012}. The study examines the differences in the microbial abundance between  samples collected from individuals residing in Malawi, Venezuela, and USA. The dataset consisted of the microbiome abundances of 100 U.S.\ individuals and 83 individuals each from Malawi and Venezuela, in addition to age and sex as the covariates.
The abundance  matrix for  $n=266$ subjects and $p=1,270$ taxa was comprised of $30\%$ zeros. 
Along with the proposed ZIBNP approach, we analyzed the data using the methods ANCOMBC, fitZIG and DESeq2. Since MaAsLin2 in its current form does not have a global significance test for more than two groups, we did not include this method in this analysis.

\begin{figure}
        \centering
        \includegraphics[height=8cm,width=8cm]{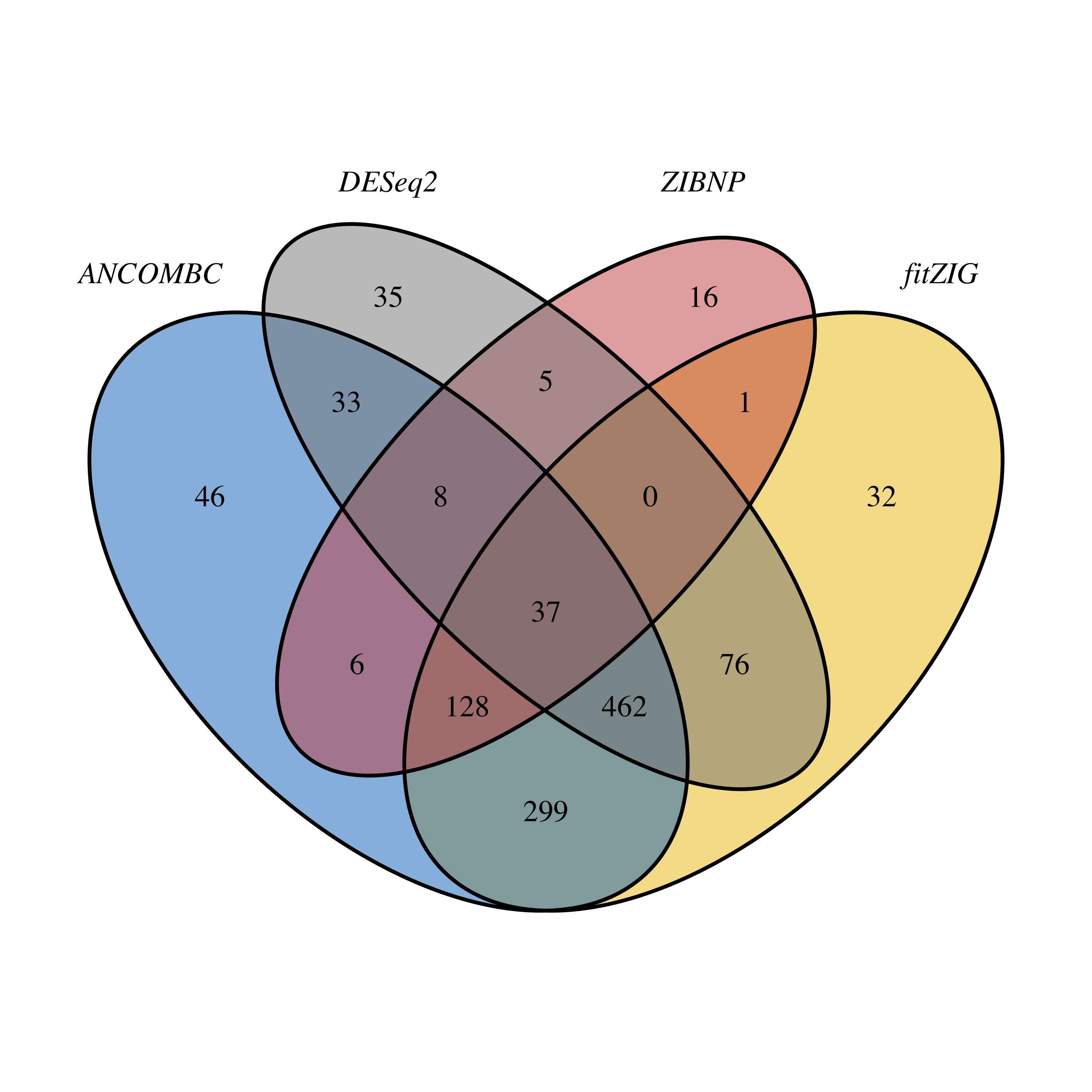}
        \caption{Venn diagram of the number of DA taxa detected by the methods ANCOMBC, DESeq2, fitZIG (MetagenomeSeq), and ZIBNP in the Global Gut dataset.}
        \label{fig:venn_globalgut}
\end{figure}

The results are shown in Table  \ref{tab:globalgut_Jaccard} and graphically summarized in Figure \ref{fig:venn_globalgut}. The  four methods detected a common set of $37$ DA taxa, as  seen in Figure \ref{fig:venn_globalgut}. ANCOMBC and fitZIG shared a large proportion of DA taxa and had a Jaccard index of $82\%$. FitZIG (MetagenomeSeq) declared $1,035$ (out of $1,270$) taxa as DA. This is consistent with the performance of FitZIG  for  the CAMP canine dataset, where it    also detected the largest number of DA taxa among all the competing methods.  

Table S1, available online as part of Supplementary Material, provides the list of $201$ DA taxa detected by ZIBNP. Among all DA taxa detected by ZIBNP, $5.5\%$ and $5\%$ of the taxa respectively belonged to the genus \textit{Prevotella} and \textit{Bacteroides}.  The original research publication \citep{Yatsunenko2012} reporting the Global Gut microbiome  results focused on pairwise comparisons, such as U.S. versus non-U.S. individuals and Venezuela versus Malawi individuals. Consequently, they are not directly comparable to  multigroup DA analysis.  However,  \cite{Yatsunenko2012} reported several taxa belonging to the genus \textit{Prevotella} as   differential for the pairwise geographical regions. Furthermore, they  stated that taxa belonging to the genus \textit{Bacteroides}  were significantly more abundant in U.S. individuals compared to non-U.S. individuals, in conformity with the findings of the proposed ZIBNP approach.

\begin{table}[]
        \centering
        \begin{tabular}{lllll}
        \hline
          & ANCOMBC  & fitZIG  &  DESeq2 & ZIBNP \\
        \hline
        ANCOMBC & 1 & 0.82 & 0.48 & 0.17 \\
        fitZIG & 0.82 & 1 & 0.51 & 0.16 \\
        DESeq2 & 0.48 & 0.51 & 1 & 0.06 \\
        ZIBNP & 0.17 & 0.16 & 0.06 & 1 \\
             \hline
        \end{tabular}
        \caption{For the Global Gut microbiome dataset, pairwise Jaccard index of the DA taxa  detected by different statistical methods.}
        \label{tab:globalgut_Jaccard}
    \end{table}

\section{Discussion}\label{S:discussion}

Differential abundance  analyses between  multiple groups of study samples  help  identify novel therapeutic targets for disease treatment. However,  few existing methods for differential analysis perform consistently well for different datasets \citep{Weiss2017}. 
Motivated by the challenges of high-dimensionality,  sparsity, and compositionality typical of  microbiome data with covariates, we propose a  novel zero-inflated Bayesian nonparametric
(ZIBNP)  model that adjusts to the distinctive data characteristics. Key contributors to our strategy's success are (i) the ability to relate the sampling depths (i.e., total counts) to inferential precision while  accommodating the inherent compositionality  of the  data, and (ii) a model-based censoring framework that learns the stochastic relationship between the
  sampling depths and  pattern of zeros to effectively impute  missing data and,  thereby, improve the  accuracy of differential   analysis. 

Latent clusters induced by the nonparametric Chinese restaurant process alleviate    collinearity issues due to the  small $n$, large $p$ microbiome datasets. The  compositional aspects are modeled by a regression framework for the log-ratios of cluster-specific group parameters. The taxa differential statuses are then evaluated as deterministic functions of 
 multivariate group-cluster  regression parameters. Through
simulation experiments  and data analyses, we demonstrate the potential of  ZIBNP  as a reliable tool  of the standard toolbox of  wide-ranging microbiome investigations. 

As suggested by several studies \citep[e.g.,][]{Sankaran2014,Xiao2017}, the accuracy of statistical methods for differential analysis can  be further improved by incorporating the   phylogenetic distances between the taxa. Looking ahead, our  research will focus on utilizing this valuable information.  Commented R
code implementing the ZIBNP approach is  available on GitHub at https://github.com/archiesach/ZIBNP.  We are  developing a faster implementation using
high-performance Rcpp subroutines that will also be publicly available on GitHub.

\bibliographystyle{unsrtnat}

\bibliography{interacttfssample}


\end{document}